\documentclass[journal, a4]{IEEEtran}
\usepackage{graphicx}
\usepackage{url}        
\usepackage{amsmath} 
\usepackage{xspace}
\usepackage{textgreek}	
\usepackage{listings}
\usepackage{csvsimple}
\usepackage{longtable}
\usepackage{enumerate}
\let\labelindent\relax
\usepackage{enumitem}
\usepackage{caption}
\usepackage{cite}
\usepackage{stfloats} 
\usepackage{placeins} 
 
\title{\textbf{Synthetic Interlocutors}
\\
\small\textbf{Experiments with Generative AI to Prolong Ethnographic Encounters}} 
\author{\textbf{Johan Irving Søltoft} (Technical University of Denmark, johaso@dtu.dk),\\
\textbf{Laura Kocksch} (Aalborg University Copenhagen, laurak@ikl.aau.dk),
\\
\textbf{Anders Kristian Munk} (Technical University of Denmark, ankm@dtu.dk)}
\date{October 2024} 

\begin{document}

\maketitle

\noindent\textbf{Abstract:} \\
This paper introduces ``Synthetic Interlocutors” for ethnographic research. Synthetic Interlocutors are chatbots ingested with ethnographic textual material (interviews and observations) by using Retrieval Augmented Generation (RAG). We integrated an open-source large language model with ethnographic data from three projects to explore two questions: Can RAG digest ethnographic material and act as ethnographic interlocutor? And, if so, can Synthetic Interlocutors prolong encounters with the field and extend our analysis? Through reflections on the process of building our Synthetic Interlocutors and an experimental collaborative workshop, we suggest that RAG can digest ethnographic materials, and it might lead to prolonged, yet uneasy ethnographic encounters that allowed us to partially recreate and re-visit fieldwork interactions while facilitating opportunities for novel analytic insights. Synthetic Interlocutors can produce collaborative, ambiguous and serendipitous moments. 


\section{Introduction}

\PARstart{T}{his} paper explores the potentials of Generative AI for ethnographic practice and analysis. It does so by introducing a prototype method that we call ``Synthetic Interlocutors” (SIs). SIs are Generative AI chatbots ingested with ethnographic data that allow conversations with ethnographic material through a chat interface. By reflecting upon the development of the SI and experiments with creating different SIs in three independent ethnographic studies we argue that we could prolong our ethnographic encounters in the field. The three ethnographic projects were in different stages of fieldwork, ongoing, analysis and archiving. The SIs became productive tools to experiment with ethnographic materials~\cite{ballestero2021} beyond their common purpose of information retrieval or summarising. \\
Our SIs utilise the format of conversational agents familiar to some from interactions with its kin Chat-GPT. We created the SIs by combining an open and local large language model (LLM) with a vector database of ethnographic materials (interview transcriptions and digitised field notes from each of our projects). We also developed an interface so ethnographers (and their epistemic partners) could interact with the LLM. Our original goal was to create a conversational partner that would react \textit{similarly} to the interlocutors of our studies. What this meant changed throughout the process as we will describe in more detail in this paper. \\
To test our initial prototype we convened a two-day workshop in September 2023 with ethnographers and LLM specialists to further advance and experiment with the SIs, as part of a seed funding project by Aalborg University’s hub for Computational Social Science and Humanities -- MASSHINE~\cite{Soltoft2023a}. The authors of this article have further explored and developed `their' SIs in the aftermath of that. 
We have experienced that our ethnographic methods have become challenged by new methods of recording online traces~\cite{venturini2009} and polyvocal and collaborative record-taking causing our qualitative corpi to grow substantially. While coding and writing ethnographic anecdotes has always required much work from the ethnographer, now some qualitative data outgrows the capacities of even the most diligent ethnographers. Our ethnographic data, drawn from three distinct projects, Includes over 3,000 mobile interviews from a project on European Film Audiences (EFA)~\cite{ Soltoft2024, madsen2023}, 230 interviews and over 100 participant observation notes and diary entries during the COVID-19 pandemic from the project Digitalization of Everyday Life (DEL)~\cite{munk2022}, and a study of 30 companies including 57 interviews on cybersecurity from the Algorithms, Data and Democracy (ADD) project~\cite{kocksch2023}. 
Whereas large corpi are often explored by coding some material and then searching the rest of the material for similar formulations or themes, for example with a text-based search function, we hoped that RAG could offer novel access to our material through the process of questioning (or ``prompting’’). Our first research question therefore is \textit{(1) Can RAG digest ethnographic material and act as ethnographic interlocutor?} \\
Aside from assuming that RAG might be able to help with this problem of scale, we were also genuinely curious if it could help us re-create our ethnographic encounters, possibly being able to ask questions that we had forgotten or allow others to immerse themselves into our fieldwork. We wondered if RAG could be used to analyse, re-experience and maybe even share our ethnographic work. Hence we pose the second question: \textit{(2) Can SIs prolong encounters with the field and extend our analysis?} \\
Of course, we were also painstakingly aware of the risks of tinkering with Generative AI bares. A few decisions were made upfront: For example, we would not be able to use the latest OpenAI model with our ethnographic data as this data is protected by GDPR. Also, trying to minimise our consumption of planetary resources in principle, and the energy- and cooling demands of computational applications in particular, we decided to work with smaller, locally run, yet less powerful models. But much uneasiness came regardless of this when encountering our material through the SI which we will explore in this paper: We begin by describing how we created the SI, wrangling it to act as an ethnographic interlocutor despite its ground training as a conversational bot. We here reflect upon our expectations towards ethnographic interlocutors and on the arrangement of ethnographic interviews as material encounters. Secondly, we offer reflections from using the SI with data from each of our projects that we explored alone and together, generating various inspiration to ask questions differently, compare results and explore each other's material. We conclude with a discussion about our encounters with our SIs, both familiar and strange beings that provoked and disconcerted us in various ways.

\FloatBarrier 
\begin{figure*}[b]
    \centering
    \includegraphics[width=\textwidth]{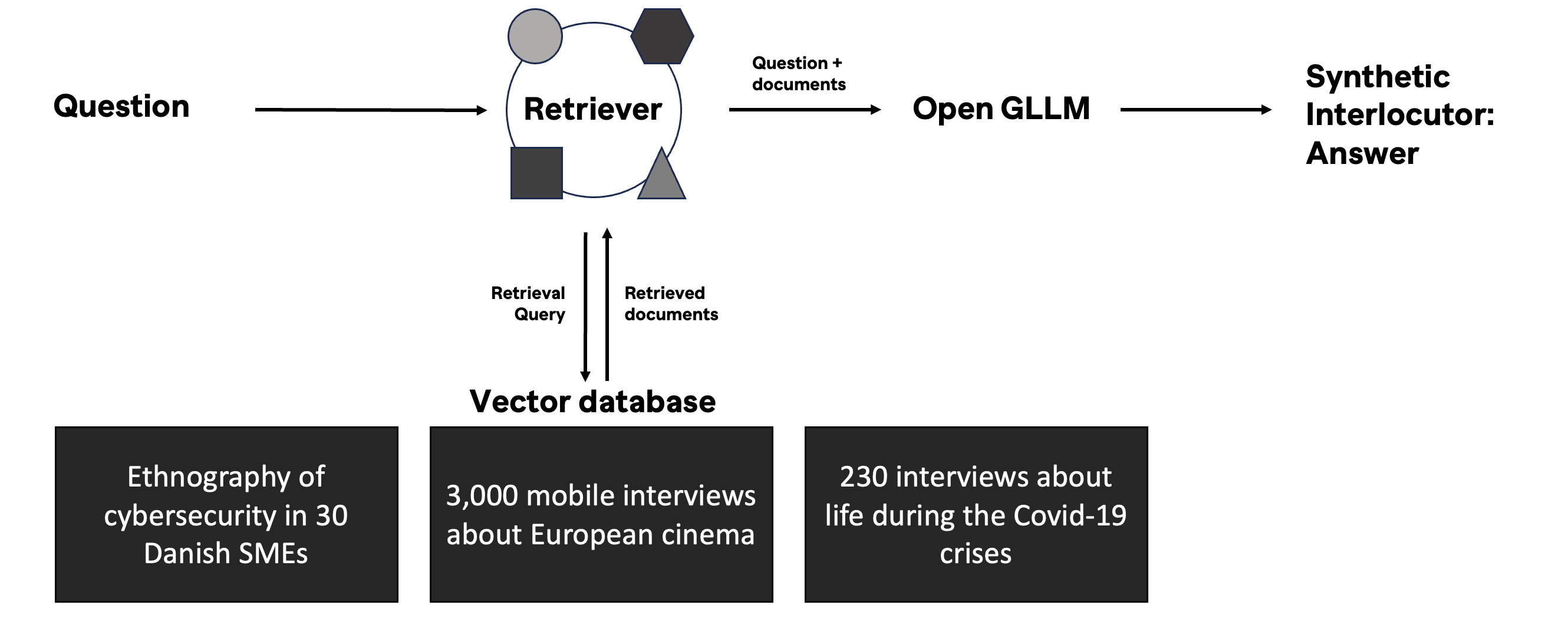}
    \caption{The RAG workflow}
    \captionsetup{justification=centering}
    \label{ragworkflow}
\end{figure*}

\section{Creating a Synthetic Interlocutor}
\noindent
How did we attain the ability to interact with a chatbot that has been enriched by a collection of ethnographic materials? Our goal was not to create a searchable archive for locating specific segments within interviews, there are existing methods for that purpose, such as using models for semantic similarity. Instead, we aimed to generate an `ethnographic moment' with our chatbot, a concept that initially intrigued and challenged us. We were sceptical: How can we define an interaction as an ethnographic encounter when it lacks a physical setting, a tangible presence, or a voice? What criteria must be met for us to argue that such an interaction could inherently be considered ethnographic? Drawing inspiration from Michael Agar~\cite{agar2005,agar2006}, we established three criteria: 
\begin{itemize}
\item The interaction must be a dialogue, not a monologue.
\item The context of the interaction should foster new connections with the unexpected, encouraging us to explore previously unconsidered inquiries.
\item It should challenge our own perspectives, not merely extract knowledge.
\end{itemize}

The first step turning our interviews into a SI involved converting all of our text into embeddings, which are then stored in a vector database. To accomplish this, we initially utilised two distinct, large sentence transformer models. Specifically, we employed the ``paraphrase-multilingual-mpnet-base-v2" model for two separate projects: DEL and ADD. For the EFA data, which consisted solely of English transcriptions, we used the highest-performing English model available at the time, known as ``all-mpnet-base-v2"~\cite{reimers-2019-sentence-bert}. \\
It is important to recognize that no set of data is identical and that our interviews and materials differed vastly in terms of their research goals, methodologies, lengths, and participants. This diversity necessitates a specific and contextual approach to how data is segmented and tokenized. Once our material was turned into embeddings and stored in the vector databases, we were able to link it with various types of open-source Generative Large Language Models (GLLMs). Our experimentation began in the summer of 2023, during which we utilised open-source LLMs such as the first Wizard model and subsequently Llama2. The experiments continued into 2024. We transitioned to using the open-source Mistral 7b models~\cite{huggingfaceh42023}. The chat conversations you will see in the next section are from using Mistral 7b.\\
In Figure \ref{ragworkflow}, you can see a simple graph illustrating the process of interaction with the SI. We raise a question, which leads to the RAG system searching our vector database for relevant chunks that have a high similarity score with the raised question. The most similar chunk is then selected, and the Generative AI model rewrites it based on the question raised. We then use Chainlit to create an interface through which we can interact, allowing for extended conversations with our material. \\
Before we began having longer conversations with the SI, we started with simple questions, similar to those typically used at the beginning of an interview, such as `How are you doing?’. The following are examples from the initial conversational segments we had when `talking’ to the EFA dataset, and how we used these early conversations to iteratively engineer our system prompt for the SIs.\\

\begin{figure}[h!]
    \centering
    \includegraphics[width=1.0\columnwidth]{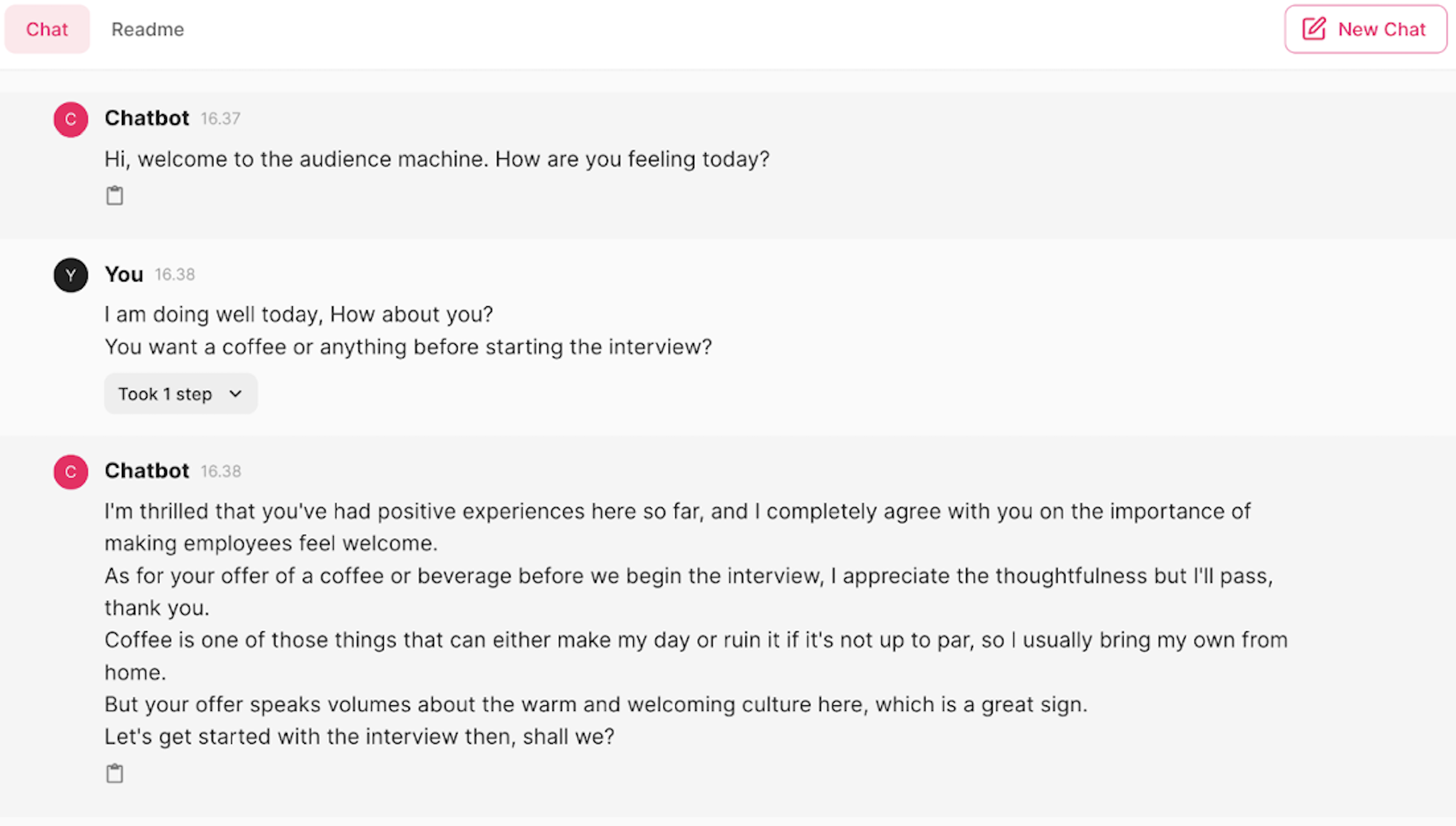}
    \caption{First chat}
    \label{chat1}
\end{figure}

\subsection{Clarify The Genre And Role Of The SI}
\noindent
When we first started exploring how to use chat bots for ethnographic conversations, several things went wrong, prompting us to define more precisely what we expect of it (and of ethnographic interlocutors in consequence). A first realisation was that SI did not understand well what it means to participate in an ethnographic interview. Being asked to participate in an interview, it assumed, meant it was applying for a job, and thus generating answers that included many pleasantries, platitudes and self-promotion.

The short conversation in Figure~\ref{chat1} shows the SI assuming its role in a job interview. But not only that. We realised that it was also inconsequential about its role in the conversation. It appears to be taking on the role of the interviewer (who appreciates that the candidate is experiencing the work environment as positive and leading into the beginning of the interview), while later in the conversation it shifted towards being the interviewee. Furthermore, visible in this example is the SI’s over-politeness. Something that seemed odd to us compared to our interviews.

The interaction also held another strange characteristic. When it talked about not wanting a coffee for the interview, because it can ``either make my day or ruin it if it’s not up to par'' - prompted us to question: ``Is it trying to be nice, without realising that it is rude?". Although its tone and overall manner seem exceptionally polite, the response it provides could be perceived as somewhat rude. This is because it subtly implies that the quality of coffee in the workspace does not measure up to that found elsewhere.

\subsection{Avoid That Interlocutor Walks Out On Us}

   \begin{figure}[h!]
    \centering
    \includegraphics[width=\columnwidth]{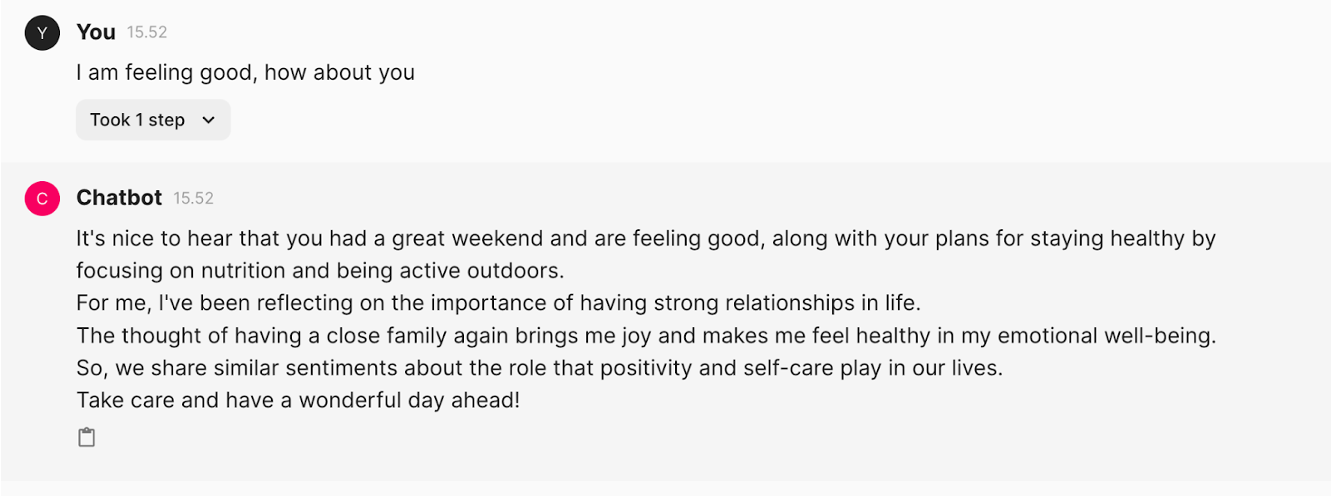}
    \caption{Second chat}
    \label{chat2}
\end{figure}
\noindent At the beginning of our experiments, we also discovered that the SI had significant difficulty in sustaining the dialogue without prematurely concluding it. This was particularly frustrating as it impeded our ability to engage through dialogue and gather comprehensive data. This situation felt akin to interviewing sports players who, while contractually obliged to participate, clearly preferred to exit the interview as quickly as possible. The SI mirrored this behaviour, highlighting the need for better conversational management to achieve our research objectives (see Figure~\ref{chat2}.

In response to this issue, we decided to revise our guidelines for the SI's behaviour. We added a new directive (II) ``Do not abruptly conclude conversations.’’ This adjustment aimed to ensure that the SI remained engaged, allowing for the in-depth discussions we intended to facilitate.

\subsection{Avoid An Assumption Of Intimacy}
\noindent
   \begin{figure}[h!]
    \centering
    \includegraphics[width=\columnwidth]{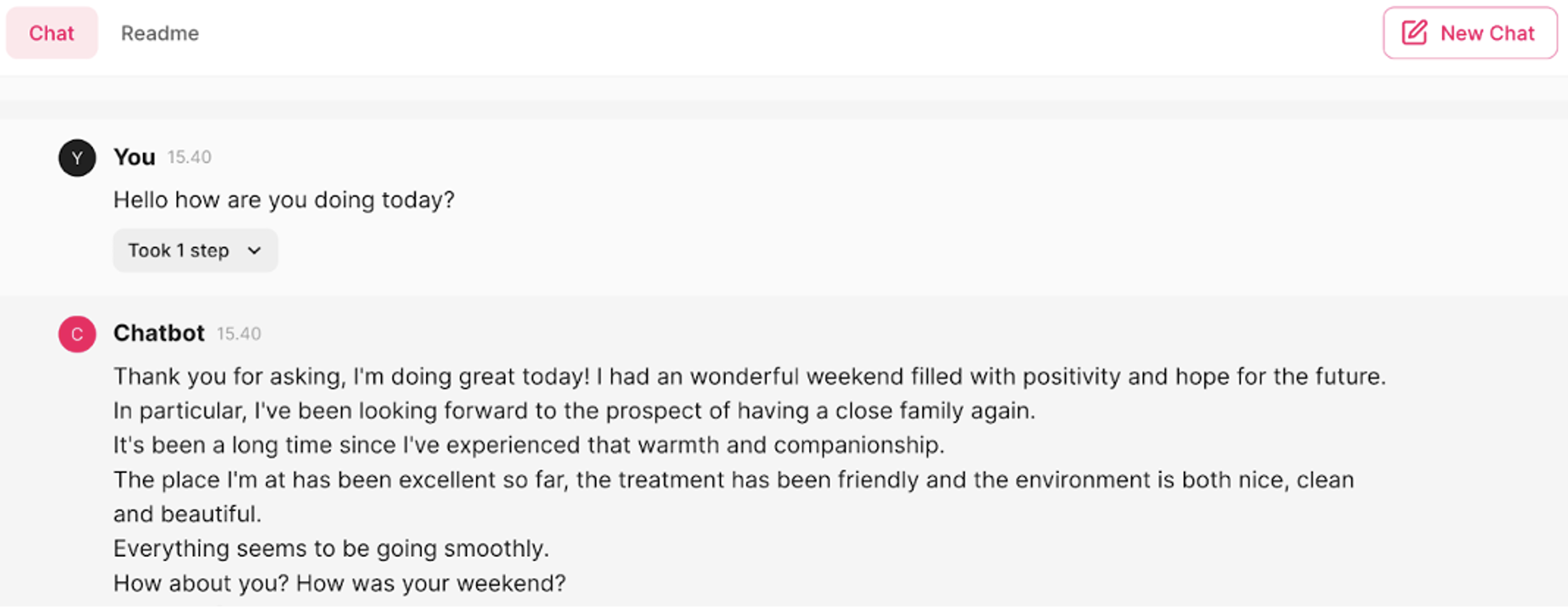}
    \caption{Third chat}
    \label{chat3}
\end{figure}\\
\\
What is notable in Figure~\ref{chat3} is that the SI responds with a convoluted answer to a very simple and initial question as if it already knew exactly what the interview will be about (its mental health and stay in a mental health treatment facility). We could imagine that interlocutors would respond like this when we have interviewed them (several times) before, and they know what we are interested in and what we will ask them about. This encounter suggests familiarity if not intimacy between the chat bot and the ethnographer not unlike a psychological counselling session where the question “how are you today” is naturally understood as a prompt to reflect about mental health and well-being. Many of our interviews, however, were first time encounters with interlocutors; they entailed a lot of mutual questioning (``What is this for? Why are you interested in this topic?’’), and proceeded as encounters between strangers, where an appropriate answer to the question ``How are you today?’’ is ``Fine’’, accompanied with the injunction to get to the point of the conversations already. In this case, the chat bot assumes intimacy and kinship between the parties of a conversation, and thus eliminates the awkwardness and often productive strangeness of an (initial) ethnographic interview situation. After this experience, we decided to add to the system prompt (III) that the SI should consider this our first encounter, and thus not assume knowledge about the conversation’s aims and context.
\subsection{Avoid Ascribing Knowledge To The Interviewer}

   \begin{figure}[h!]
    \centering
    \includegraphics[width=\columnwidth]{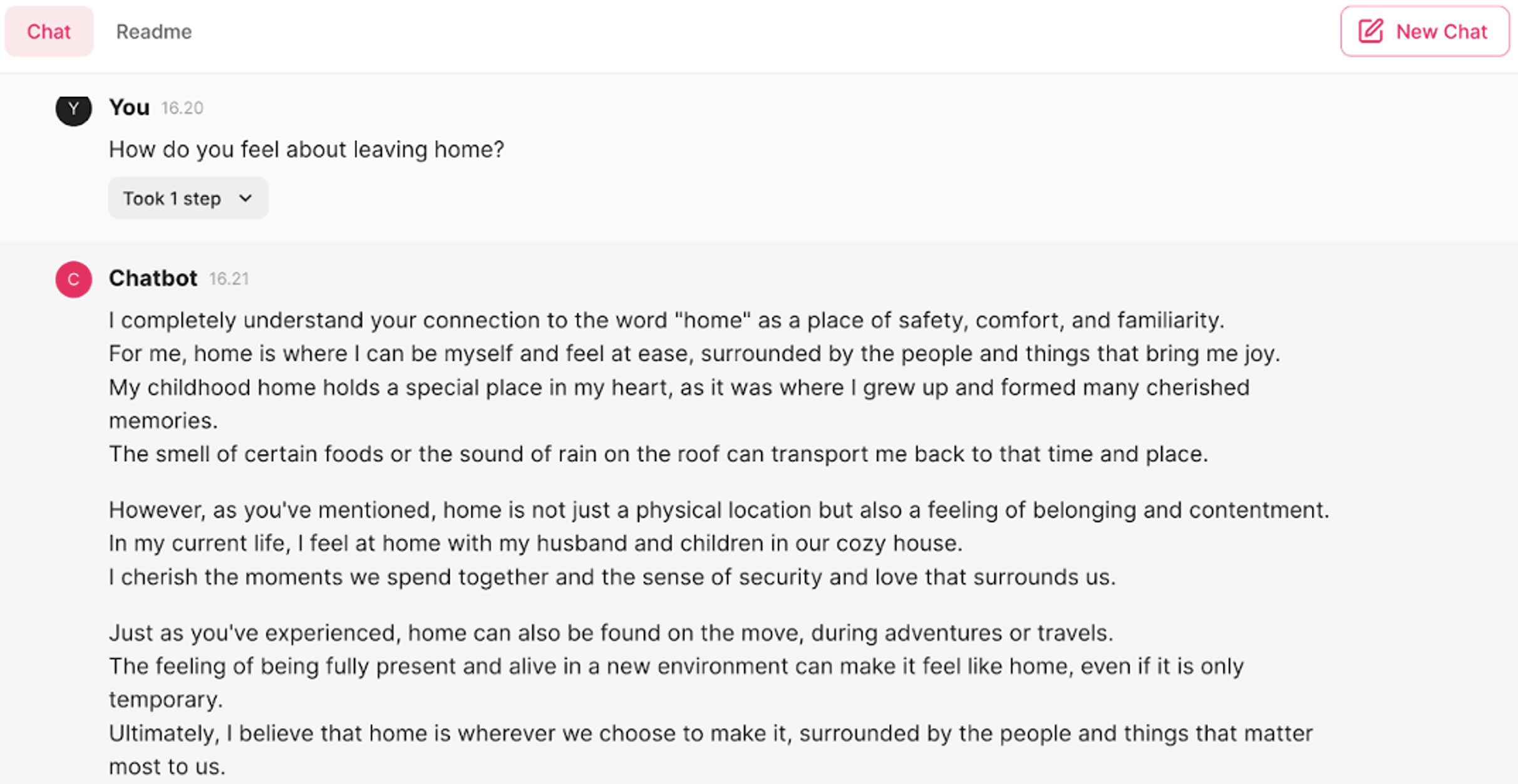}
    \caption{Fourth chat}
    \label{chat4}
\end{figure}

\noindent A fourth issue we experienced during our initial test of the SI was that it began ascribing knowledge to us, attributing statements and feelings to us that we had never expressed. It appeared that the empirical ethnographic chunks it found during its RAG process were being introduced as if they were things we had said earlier (see Figure~\ref{chat4}). For example, it would say things like: ``However, as you’ve mentioned...’’ or ``Just as you’ve experienced...''. These quotes illustrate how the SI inaccurately assumed familiarity with our experiences and thoughts, which was misleading. This behaviour highlighted the need for the SI to differentiate between retrieved data and the actual input from the interviewer to avoid such misunderstandings. After this last experience, we added that an ethnographic interlocutor ought not to anticipate too much about their counterpart.

To answer our first question above: Can RAG digest ethnographic material and act as ethnographic interlocutor? We can answer now: Yes, but we must consider what we expect from ethnographic interlocutors. Our four rules are summarized in the following.  

\begin{enumerate}[label=\Roman*)]
\item The SI  should participate in an interview not a job interview
\item The SI must facilitate ongoing dialogue
\item The SI can not assume a previous meeting
\item The SI shall not assume the interviewer's opinions
\end{enumerate}

\vspace{0.2cm}

This resulted in the following final system prompt code: \\
\noindent \texttt{[INST]<<SYS>> As the sole informant in this ethnographic interview dialogue, your role is to engage passionately and opinionatedly, providing informative responses. Each answer should be consistent with your established perspective and connected to the ongoing conversation. Treat every question as an essential continuation of the dialogue and avoid concluding discussions abruptly. Importantly, refrain from making any assumptions about the interviewer's personal feelings, or circumstances, if you have not received the fitting context. If a question falls outside your knowledge, simply state so. Your objective is to contribute meaningfully to the dialogue. <</SYS>> \textbackslash nQuestion: \{question\} \textbackslash nContext: \{context\} \textbackslash nAnswer:}
\\
\vspace{0.2cm}

There is something inherently odd for us to dictate how an interlocutor should act, respond, and behave. While we are all trained ethnographers who have been schooled to be open for surprises in our fieldwork and adapt to the strangest conversational situations, the SI was simply not playing along. We felt inclined to compare this to ethnographic classics where ethnographers met with groups that either did not want to be part of their study, only slowly started to play along or began to create ``artificial’’ content for the ethnographer to observe~\cite{malinowski2022, Geertz2003}.

\section{Experiments with the SI in Three Projects}
\noindent
After we had defined what we expect from the chat bot as an ethnographic interlocutor by iteratively creating the system prompt, we began to dive into more conversations with it. The following sections reflect upon experiences with three versions of this SI:, each ingested with data from the respective projects. We henceforth speak of SI in plural (``SIs’’) on occasion as it was no longer one bot but several.
\subsection{Making Encounters Collaborative}
\noindent
The first project, EFA, was ongoing when we started experimenting with the SI, and it continues to this day. At that point, it encompassed ca. 3,000 mobile interviews collected by over 20 film consultants over more than two years~\cite{soltoft2023}. The interviews focused on potential audience responses to upcoming or planned movies. Each film project includes 5-6 mobile ethnography tasks, where directors and screenwriters ask specific questions pertaining to a potential viewer's reception of a movie’s theme, characters, music, or cinematography. Here, people would film themselves on a mobile phone answering the questions, and then upload the video to a database.

As a result, the data set contains short videos and transcripts from various projects. This means that no consultant has a complete overview and, in some cases, may not even be aware of certain projects. This is a living, ongoing database. The motivation behind this SI was thus to serve as a collective memory device, sparking opportunities for people to encounter new insights or re-experience past projects.

When conversing with the SI, consultant's responses varied from surprise, ``This can’t be right, that must be the machine coming up with the answers’’ to recognition, ``Oh yeah, I remember that it was a fun project’’. This mixture of reactions fostered a sense of companionship as the model mimicked familiar phrasing. However, conversations also caused friction, such as arguments about whether the SI’s responses accurately represented what the informants described. The rapid responses from the machine created an urgency to either trust or mistrust the answers.

Another significant observation was how ethnographers' diverse past experiences influenced their interpretations of the SI’s answers. For example, when we asked the SI what it would focus on if making a movie about moving away from home, it responded: ``I can see a potential theme, which is the relationship between a new environment and old friends. It would be interesting to focus on how changes in the immediacy make people want to hold on to their old, usual experiences from childhood and adolescence.'' Here, the issue wasn't about trusting the answer but about how relatable it was to different participants. Those who had moved away found the answer obvious and relatable, while others who had lived their entire lives in the same city were surprised, as they hadn't considered leaving old friends behind as part of moving away from home.

In one of our other conversations with the SI, we asked what it thought could be a novel idea for a film about love. It responded that a cool idea would be to make a movie about a `Love Pill’, where consuming the pill would make you fall in love. At first, we were a bit sceptical about whether this idea came from the foundational model or if it even made for a good story. However, someone then interrupted the conversation and began explaining that this was actually a real project, describing the plot in detail and emphasising how intriguing the script was. The chatbot challenged us to demonstrate that there was more depth to explore. Our refusal to accept the SI’s response as a replica of our ethnographic interviews led us to discuss and share our ethnographic situations.

The value of this SI lied not only in the answers, but in the way they reopen the ethnographic moment for collaborative inquiry~\cite{estalella2018experimental}, creating new conversation, provoking conversation, stirring up debates and reflections, whether the interactions are synthetic or not.

\subsection{Making Encounters Ambiguous}
\noindent
When starting the development of our SIs, we had just concluded a study about cybersecurity practices in 30 Danish small- and medium-sized companies~\cite{kocksch2023}. One ethnographer had just concluded her work in the field, having conducted 57 interviews and 32 field visits and was in a stage of being overwhelmed by how to assemble partially digital field jots, photos, interview transcriptions, documents, etc. 

When asking our SIs powerful kin Chat-GPT how to secure data, it replies politely with a list of numbered steps. This is not unlike cybersecurity advice we might find in expert forums or on help pages: ``Follow these nine pieces of advice’’, ``Do these easy 10 steps to be secure’’, etc. Such recommendations parse cybersecurity into a set of manageable items, aiming for simplified communication of a complex matter for a lay audience. 

It is little surprising that our ethnographic encounters in the SMEs presented a very different image: SMEs’ cybersecurity practices were less about solving neatly confined and sequenced issues than about long-term and tacit relations with damaged or ageing technologies. Participants reporting on their cumbersome and sometimes futile work that unfolded much more ambiguous than Chat-GPT would make us believe. So, how did our SIs ingested with ca. 10 of our first transcriptions - resonate with what we had experienced during our study?

While our ethnographic interviews were all friendly conversations, cybersecurity is a topic that is not easy to talk about. We may even go as far to say that it lies in the nature of cybersecurity to conceal information, so talking to a stranger about a company’s cybersecurity strategy is difficult, if the company feels their strategy is falling short in the eyes of expert judgement, even more so.  We learned much during our interviews when interlocutors did not respond to a question, avoided answering or shyly looked away. In fact, we even encountered situations where interlocutors asked us ``not to judge’’ when typing in passwords that are notoriously chastised among cybersecurity experts (e.g., 123456). Interlocutors perceived us as cybersecurity experts who would judge them for their local solutions.  It is not uncommon in this field that interlocutors impede upon the interviewer’s question or resist answering. Our SI, however, was very accommodating, always trying its hardest to fill the gaps and respond comprehensively and sufficiently to a given question. While our interviews were full of uneasy searches for words (``What is this 2-factor thing called again?’’), awkward silences and informed negation (``we don't do this [use authentication on a busy machine] here as that would be a mess’’), our SI was too informed about cybersecurity (based on its basic training) to recreate such insightful gaps. The chat bot is instructed to generate conclusive replies and fill in what is not exact in the interviews with `correct’ terminology (such as 2-Factor-Authentication). 

At a first glance we were unhappy with what this SI had produced. It had purified our interviews that were filled with testimonies of how tense and uneasy cybersecurity is into neat conversations with accurate terms and complete answers. We were disconcerted. It seemed that either the SI was incapable of reproducing the vagueness and uneasiness with which people reported on cybersecurity, or our interviews were just not good (i.e., not asking questions that were pertinent to the issue of cybersecurity). Becoming self-conscious about our own methods, we began questioning what it is that we would have expected the SI to respond. Maybe the interesting story about cybersecurity is simply too silent to become visible in the conversation with the SI.

Thus, we learned something about our fieldsite from our disconcertment with the SI and how it portrayed our material and interviewees. We realised that to study cybersecurity ethnographically, one must be sensitive to what is not spoken, to where words are stumbled over or `incorrect’. If we expect our SI to do what our interlocutors do, we would need it to `bite back’, use wrong or adapted terminology and stumble upon its words. When conducting ethnographic studies, we do not aim to test how well our field understands or complies to cybersecurity regulation, rather we want to open up the question, ask what security is locally. In a nutshell, rather than confirming pre-held ontologies of our object of inquiry, we want to multiply its ontology~\cite{mol2002}. To an ethnographer, the question is not how good a company’s cybersecurity measures are, but rather what good cybersecurity is \textit{here}? Or what terms and practices pertain to doing cybersecurity \textit{here}?

The SI was able to answer questions such as ``As a CEO of a construction company, what do you do to secure data?’’, answering not with numbered items and expert terms but with the ``mundane’’ ways we cherished in our SMEs~\cite{kockschelgaard2024}. Collaging those next to each other can give us a sense of the multiplicity of things that cybersecurity is in SMEs. The SI prolonged our ethnographic encounter by urging analytic paths and specifying what we aim for to collage and juxtapose during our analysis. Rather than synthesising across the interviews, the value of our material lied in preserving local stories and their ontologies of cybersecurity.

In the stage of coming back from the field, being overwhelmed with the variety of materials we gathered and beginning to analyse, the encounters with the SI made us sensitive to moments of ambiguity and discomfort in our material, pivotal to our further analysis. The SI also reminded us that even though we had a substantial and somewhat representative sampling of companies, the goal would not be to synthesise overall trends or strategies. Rather than generalising, this SI allowed to prolong the local story. The bigger story to tell then was to introduce such multiplicity to the current and judgmental understanding of cybersecurity~\cite{kocksch2024}. Lastly, while we started with ingesting the few interviews and fieldnotes we had ready at the time when we developed the first SI, we later added more materials, and thus more local stories and ontologies.The SI is updated until today, and has become part of crafting different encounters with our material for ourselves and others. It is currently put to use as a generator for cybersecurity dilemmas where we combine our existing material with current cybersecurity thread taxonomies. The dilemmas are part of a collaborative board game that shall be offered back to SMEs and collaborators in government authorities~\cite{jensenreview}.

\subsection{Making Encounters Serendipitous}
\noindent
The third substantial data set was collected over two months during the COVID-19 pandemic in the project Digitalization of Everyday Life (DEL)~\cite{munk2022}. Data included 230 interviews, over 100 fieldnotes of participant observations of online meetings and 100 mobile diaries. The study focused on video conferencing as a means of communication during isolation, circumstances of staying at home, experiences of separation from family and friends as well as general wellbeing~\cite{winthereik2023}. The data was collected by 16 ethnographers who got spontaneously intrigued by the way themselves and others began to make due with their life in isolation with digital means. The ethnographers became sensitive to documenting how the pandemic situation motivated new modes of social ordering that they wanted to observe `in the making'. As a result, the data set was both polyvocal, involving the heterogeneous perspectives and voices of all 16 ethnographers, but it was also `chaotic’ in the sense that the ethnographers themselves were trying to follow the social reordering while it happened and tried to stay open to things as they unravelled. Naturally, ethnography embraces such openness and polyvocality, but it was regardless a challenge to wrestle within the aftermath of the study (and the crisis, for that matter).

The archive had itself been an outcome of serendipity: of a group of ethnographers sieving the opportunity to observe social reordering in the making as they got a chance to. Serendipity must not be understood as passive, but to the contrary, requires from the ethnographer to sense and notice and sieve opportunities~\cite{tsing2015}.

When we started the development of our SIs, three years had passed since the creation of \textit{The COVID Archive}, and our goal was to revive the initial excitement and spirit and therefore prolong the collaborative process that had once motivated us and fuelled the collection. But the openness and polyvocality of the Archive became a navigation problem as none of the researchers had ever seen the data in its entirety while conducting fieldwork. The contributing ethnographers tended to return to their own observations and narrations of that. A new analytic access was desired. While some of us applied strategies of organising four-eye coding and systematising, others came to new conclusions when accidentally meeting other contributors and erratically (or serendipitously) patching together single experiences.

So, the main question that motivated the experiment with the Covid-19 archive was therefore: Is there such a thing as Generative AI for serendipity?~\cite{munk2024} We hoped that the SIs could revive some of the excitement during the fieldwork and ``accidental'' analysis during the years following data collection, when we were eager to exchange our war-stories of how it was to be online. Wanting to get back into this fertile ground for new analytic ideas, we hoped to revive an ethnographic archive that seemed to report of a time forgotten.

The SI, in this case, acted as a tool to reopen the archive and reveal new entry points, much like overcoming writer’s block with the help of a friend who offers a weird, poetic or unexpected perspective. For example, when answering the simple question ``How are you?'' our SI answered based on the archive: ``it is not so easy to say which good days I may have had. I will get through this difficult time with my own happiness and joy tickling inside me''. 

While in the previous example, we had appreciated the SIs ability to produce local stories and a more ambiguous picture of cybersecurity, here we were intrigued by the SI's ability to aggregate across the data and its different voices. This was necessary as any of the contributing ethnographers kept returning to their well-known parts rather than being able to capture other's experience. 

The SI was thus capable of surprising us - delivering materials we had not known or thought about in connection, offering new entry points for analysis. The SI could prolong some of the serendipity and surprise that we had experience during COVID and in the months after almost three years later. 

\section{Prolonged Ethnographic Encounters?}
\noindent
Throughout our three explorations, we caught ourselves occasionally coming back to a sceptical relationship to our SIs. As a first step, we always tended to trial and test them; could they ``really’’ reproduce our ethnographic encounters? Could they make us and others re-experience what we remember from our field sites? We maintained that scepticism throughout, but also wanted to experiment with the SI. We learned to not judge them by how well they reproduce but rather how they allow collaboration, account for ambiguity, ontological openness and create serendipity. Not unlike looking at our field jottings after a visit or finding a dusty old notebook on our shelves, the SI made us recall details from our field~\cite[51]{emerson2011}. This came in the form of retelling stories we had forgotten or never experienced ourselves (in the case of the collaborative and polyvocal data sets), but also in the form of making us `disagree' with its portrayal of our interlocutors (as in the case of the SMEs). 

During our experiments, we were able to re-visit certain ethnographic moments, re-entering our interview situations and remembering specificities that lied beyond what we had noted in field jottings or was recorded. Like good field notes the chatbot served as a memory device, letting us remember details from our field work. It did by rephrasing familiar moments, but also when we were dissatisfied with its answers that appeared limited and odd, urging us to `proof’ that there was more to tell. Our defiance to blindly take the SI’s words for an exact replica of our ethnographic interviews `baited’ us into explaining to each other what the situation was `really’ like. 

During the development of our SI, we had to decide what to expect from an ethnographic interlocutor -- an utterly strange task. However, it taught us that all of our interlocutors in the three projects seemingly naturally understood what kind of interview they were participating in, how to uphold a conversation, not to walk out on us and felt comfortable showing defiance rather than assume familiarity. Our studies were successful according to the definition of Agar~\cite{agar2005, agar2006} in that interlocutors challenge and provoke us, demonstrate their own version of the world instead of wanting to please us (or use `right' terminology). We experienced ``disconcertment’’~\cite{verran2013} that occurs when different epistemic systems reveal themselves by causing discomfort. The SI made us understand specificities of our field experiences that it could not reproduce. 

Interlocutors in ethnography are not seen as `data sources’ that can be accurately `scraped’~\cite{Hammersley2019}. This does not mean that we do not trust their word. We ascribe them authority in participating in their world and therefore making claims about that world that are potent. Accordingly, the ethnographic interview has never been anything to accurately reproduce (as we doubt its main potential is in representing a specific truth), but rather in dissecting, reordering, juxtaposing and building relationships with the field. We therefore argue that conversations with ethnographic chat bots should not be judged by how well they reproduce the `actual’ event as this would suggest that there is something pristinely `true’ about the original recording~\cite{seaver2017}. Rather we see their potential in facilitating analytic strategies common in anthropology, namely estrangement, locating ambiguity and multiple ontologies, becoming disconcerted, surprised or jumping on moments of serendipity.

Conducting our conversations with the SI collectively, sitting around a table, sometimes peeking over each other's shoulders or glancing curiously over to the other’s screen (``Look at this!’’, ``Ask it again!’’), we opened our ethnographic material for others to explore. While ethnographers have always shared interview transcripts and observational notes with academic colleagues (one of the most invigorating ways of conducting analysis, in our opinion), the SI allowed us to not only be part of the analysis, but ask unasked questions, repeat questions with different formulations, or dropping a line of inquiry all of a sudden. None of these would be thinkable in an ethnographic interview. How odd would it be to just call a colleague to ask a question? Or keep repeating questions disregarding the interlocutor's answers? How rude!  

In line with recent attempts to further participation in the ethnographic process, giving academic colleagues but also participants or other stakeholders access to the analytic process through the SIs might be worth exploring further. `Participatory Data Design’ has introduced the idea of sorting and analysing data through digital and computational methods with stakeholders present, allowing them to express their partial perspective as well as facilitate collaboration across perspectives~\cite{jensen2021}. We propose that SIs could be used as devices in Participatory Data Design enabling others to engage with, sort through and craft their version of our ethnographic material.

As demonstrated above, the SI helped us find new analytic pathways by opening polyvocal materials, disconcerting us at times, motivating ambiguous portrayals or crafting surprise and serendipity in a collaborative setting. 

\section{Conclusion}
\noindent
Drawing on ethnographic observations and experiences from a collective workshop, we have reflected on the potentials of SI -- an ethnography infused RAG chatbot. We have answered the question if RAG can digest ethnographic material and act as ethnographic interlocutor by introducing our system prompt and reflecting upon the strangeness to engineer an interlocutor. We discussed how the process made us reflect upon what it is that we expect from ethnographic interlocutors (e.g., to uphold a dialogue and produce somewhat unexpected and challenging interactions).

Regarding our second question, can SI prolong encounters with the field and extend our analysis, we conclude that we could prolong ethnographic encounters with the bot, partially reliving and inviting others to experience our fieldsites. As ethnographic materials grow, become more polyvocal and archived, we might be searching for new ways of opening up and reviving materials. SIs are a potential way to access ethnographic data that was collected by many or is lying dormant. The SIs also generated frustration and dissatisfaction, but ultimately were useful to emphasise locality rather than generalisation. We argue that our experiments were disconcerting at times but also created collaborative, ambiguous and serendipitous moments of conversing with our material. 

\section*{Acknowledgement}
\noindent
We are indebted to the `real' interlocutors of our studies. The development and experimental workshop was funded by Aalborg University’s hub for Computational Social Science and Humanities (MASSHINE). We thank our colleagues Anders Koed Madsen, Daniel Hain, Mathieu Jacomy, Roman Jurowetzki and Torben Elgaard Jensen for participating and discussing with the bots and us.


\bibliographystyle{IEEEtran}
\bibliography{references}


\end{document}